\documentclass{aa}
\usepackage{mathtools}
\usepackage{multirow}
\usepackage{booktabs}
\usepackage{graphicx}
\graphicspath{{Fig/}}
\usepackage{menukeys}
\usepackage{algorithm}
\usepackage{ulem}
\usepackage[noend]{algpseudocode}
\def\folio{\ifnum\pageno=1\nopagenumbers\else\number\pageno\fi}

\def\lax    {\ifmmode{_<\atop^{\sim}}\else{${_<\atop^{\sim}}$}\fi}
\def\gax    {\ifmmode{_>\atop^{\sim}}\else{${_>\atop^{\sim}}$}\fi}
\newbox\grsign      \setbox\grsign=\hbox{$>$} 
\newdimen\grdimen   \grdimen=\ht\grsign
\newbox\simgreatbox \setbox\simgreatbox=\hbox{\raise.5ex\hbox{$>$}\llap
                        {\lower.5ex\hbox{$\sim$}}}\ht1=\grdimen\dp1=0pt
\newbox\simlessbox  \setbox\simlessbox =\hbox{\raise.5ex\hbox{$<$}\llap
                        {\lower.5ex\hbox{$\sim$}}}\ht2=\grdimen\dp2=0pt

\newbox\grsign \setbox\grsign=\hbox{$>$} \newdimen\grdimen \grdimen=\ht\grsign
\newbox\laxbox \newbox\gaxbox
\setbox\gaxbox=\hbox{\raise.5ex\hbox{$>$}\llap
     {\lower.5ex\hbox{$\sim$}}}\ht1=\grdimen\dp1=0pt
\setbox\laxbox=\hbox{\raise.5ex\hbox{$<$}\llap
     {\lower.5ex\hbox{$\sim$}}}\ht2=\grdimen\dp2=0pt
\def\gax{\mathrel{\copy\gaxbox}}
\def\lax{\mathrel{\copy\laxbox}}
\def\boxit#1    {\vbox{\hrule\hbox{\vrule\kern3pt
                  \vbox{\kern3pt#1\kern3pt}\kern3pt\vrule}\hrule}}
\def\h      {\ifmmode{^{\rm h}}\else{$^{\rm h}$}\fi}
\def\m      {\ifmmode{^{\rm m}}\else{$^{\rm m}$}\fi}
\def\s      {\ifmmode{^{\rm s}}\else{$^{\rm s}$}\fi}
\def\decas    {\ifmmode{{\rlap.}{''}}\else{${\rlap.}{''}$}\fi}
\def\mum     {\ifmmode{\mu{\rm m}}\else{$\mu{\rm m}$}\fi}
\def\s      {\ifmmode{^{\rm s}}\else{$^{\rm s}$}\fi}
\def\deg      {\ifmmode{^{\circ}}\else{$^{\circ}$}\fi}
\def\as     {\ifmmode {\rlap.}$\,$''$\,$\! \else ${\rlap.}$\,$''$\,$\!$\fi}
\def\decsec  {\ifmmode {\rlap.}$\,$^{\rm s}$\,$\! \else ${\rlap.}$\,$^{\rm s}$\,$\!$\fi}\def\decs  {\ifmmode {\rlap.}$\,$^{\rm s}$\,$\! \else ${\rlap.}$\,$^{\rm s}$\,$\!$\fi}

\def\kms    {\ifmmode{{\rm km~s}^{-1}}\else{km~s$^{-1}$}\fi}

\def\Mspy   {\ifmmode {M_{\odot} {\rm yr}^{-1}} \else $M_{\odot}$~yr$^{-1}$\fi}
\def\Mdot   {\ifmmode {\dot M} \else $\dot M$\fi}
\def\mhd    {\ifmmode {n_{{\rm H}_2}} \else $n_{{\rm H}_2}$\fi}
\def\mhcd   {\ifmmode {N_{{\rm H}_2}} \else $N_{{\rm H}_2}$\fi}

\def\El      {\ifmmode{E_{\ell}}\else{$E_{\ell}$}\fi}
\def\beam    {\ifmmode{\theta_{\rm B}}\else{$\theta_{\rm B}$}\fi}
\def\mjyb   {\ifmmode {{\rm mJy~beam}^{-1}} \else{mJy~beam$^{-1}$}\fi}
\def\mujyb   {\ifmmode {\mu{\rm Jy~beam}^{-1}} \else{$\mu$Jy~beam$^{-1}$}\fi}
\def\Trot   {\ifmmode{T_{\rm rot}}\else$T_{\rm rot}$\fi}    
\def\Tvib   {\ifmmode{T_{\rm vib}}\else$T_{\rm vib}$\fi}    
    
\def\Teff   {\ifmmode{T_{\rm eff}}\else$T_{\rm eff}$\fi}

\def\ITRS   {\ifmmode{\smallint {\rm T}_{R}^{*}dv}\else{$\smallint 
{\rm T}_{R}^{*}dv$}\fi}
\def\ITRS   {\ifmmode{\smallint {\rm T}_{R}^{*}dv}\else{$\smallint 
{\rm T}_{R}^{*}dv$}\fi}
\def\ITAS   {\ifmmode{\smallint {\rm T}_{A}^{*}dv}\else{$\smallint 
{\rm T}_{A}^{*}dv$}\fi}

\def\lefttitle#1  {\noindent \hangindent=18.0pt \hangafter=1 {#1} \par}
\def\vol#1  {{\bf {#1}{\rm,}\ }}
\font\tenssb=cmssbx10
\font\tenbf=cmbx10
\font\sevenbf=cmbx8
\font\fivebf=cmbx6
\def\unetdemi    {\smallskipamount=6pt plus2pt minus2pt
                  \medskipamount=12pt plus4pt minus4pt
                  \bigskipamount=24pt plus8pt minus8pt
                  \normalbaselineskip=16pt plus0pt minus0pt
                  \normallineskip=2pt
                  \normallineskiplimit=0pt
                  \jot=6pt
                  {\def\smallskip {\vskip\smallskipamount}}
                  {\def\medskip   {\vskip\medskipamount}}
                  {\def\bigskip   {\vskip\bigskipamount}}
                  {\setbox\strutbox=\hbox{\vrule 
                    height17.0pt depth7.0pt width 0pt}}
                  \parskip 12.0pt
                  \normalbaselines}
\def\smallerspace {\smallskipamount=3pt plus0pt minus0pt
                  \medskipamount=6pt plus0pt minus0pt
                  \bigskipamount=10.5pt plus0pt minus0pt
                  \normalbaselineskip=10.5pt plus0pt minus0pt
                  \normallineskip=1pt
                  \normallineskiplimit=0pt
                  \jot=3pt
                  {\def\smallskip {\vskip\smallskipamount}}
                  {\def\medskip   {\vskip\medskipamount}}
                  {\def\bigskip   {\vskip\bigskipamount}}
                  {\setbox\strutbox=\hbox{\vrule 
                    height8.5pt depth3.5pt width 0pt}}
                  \parskip 0pt
                  \normalbaselines}
\def\memospace    {\smallskipamount=4pt plus1pt minus1pt
                  \medskipamount=6pt plus2pt minus2pt
                  \bigskipamount=14pt plus6pt minus6pt
                  \normalbaselineskip=14pt plus0pt minus0pt
                  \normallineskip=1pt
                  \normallineskiplimit=0pt
                  \jot=4pt
                  {\def\smallskip {\vskip\smallskipamount}}
                  {\def\medskip   {\vskip\medskipamount}}
                  {\def\bigskip   {\vskip\bigskipamount}}
                  {\setbox\strutbox=\hbox{\vrule 
                    height17.0pt depth7.0pt width 0pt}}
                  \parskip 2.0pt
                  \normalbaselines}
\def\memowidespace    {\smallskipamount=5pt plus1pt minus1pt
                  \medskipamount=7.5pt plus2pt minus2pt
                  \bigskipamount=17.5pt plus6pt minus6pt
                  \normalbaselineskip=17.0pt plus0pt minus0pt
                  \normallineskip=1.25pt
                  \normallineskiplimit=0pt
                  \jot=5pt
                  {\def\smallskip {\vskip\smallskipamount}}
                  {\def\medskip   {\vskip\medskipamount}}
                  {\def\bigskip   {\vskip\bigskipamount}}
                  {\setbox\strutbox=\hbox{\vrule 
                    height21.25pt depth8.75pt width 0pt}}
                  \parskip 2.5pt
                  \normalbaselines}
\usepackage[colorlinks=true, allcolors=blue]{hyperref}
\usepackage{txfonts}

\newcommand{\ttch}{$^{13}$CH }
\newcommand{\cratio}{$^{12}$C/$^{13}$C } 
\usepackage{cuted}
\makeatletter
\renewcommand*\aa@pageof{, page \thepage{} of \pageref*{LastPage}}
\makeatother
\bibpunct{(}{)}{;}{a}{}{,}

\begin{document} 

   \title{First detection of $^{13}$CH in the interstellar medium}

   \author{Arshia M. Jacob\inst{1}
          \and
          Karl M. Menten\inst{1}
          \and
          Helmut Wiesemeyer\inst{1}
          \and 
          Rolf G\"{u}sten\inst{1}
          \and 
          Friedrich Wyrowski\inst{1}
          \and
          Bernd Klein\inst{1,2}
          }

   \institute{Max-Planck-Institut f\"{u}r Radioastronomie, Auf dem H\"{u}gel 69, 53121 Bonn, Germany
   \and 
   University of Applied Sciences Bonn-Rhein-Sieg, Grantham-Allee 20, 53757 Sankt Augustin, Germany\\
   \email{ajacob@mpifr-bonn.mpg.de}}

   \date{Received September 15, 1996; accepted March 16, 1997}
  \titlerunning{First detection of $^{13}$CH in the interstellar medium}
   \authorrunning{A. Jacob et al.}

\abstract{In recent years, a plethora of high spectral resolution observations of  sub-millimetre and far-infrared transitions of methylidene (CH), conducted with Herschel and SOFIA, have demonstrated this radical to be a valuable proxy for molecular hydrogen, that can be used for characterising molecular gas within the interstellar medium on a Galactic scale, including the CO-dark component. Here we report the discovery of the ${}^{13}$CH isotopologue in the interstellar medium using the upGREAT receiver on board SOFIA. We have detected the three hyperfine structure components of the $\approx 2~$THz frequency transition from its X$^{2}\Pi_{1/2}$ ground-state toward the high-mass star-forming regions Sgr~B2(M), G34.26+0.15, W49(N) and W51E and determine \ttch column densities. The ubiquity of molecules containing carbon in the interstellar medium has turned the determination of the ratio between the abundances of carbon's two stable isotopes, \cratio, into a cornerstone for Galactic chemical evolution studies. Whilst displaying a rising gradient with Galactocentric distance, this ratio, when measured using observations of different molecules (CO, H$_{2}$CO, and others) shows systematic variations depending on the tracer used. These observed inconsistencies may arise from optical depth effects, chemical fractionation or isotope-selective photo-dissociation. Formed from C$^+$ either via UV-driven or turbulence-driven chemistry, CH reflects the fractionation of C$^+$, and does not show any significant fractionation effects unlike other molecules previously used to determine the \cratio isotopic ratio which make it an ideal tracer for the \cratio ratio throughout the Galaxy.  Therefore, by comparing the derived column densities of \ttch with previously obtained SOFIA data of the corresponding transitions of the main isotopologue $^{12}$CH, we derive \cratio isotopic ratios toward Sgr B2(M), G34.26+0.15, W49(N) and W51E. Adding our values derived from $^{12/13}$CH to previous calculations of the Galactic isotopic gradient we derive a revised value of \cratio = $5.85(0.50)R_{\text{GC}} + 15.03(3.40)$.}

\maketitle
%

\section{Introduction}

The methylidene radical, CH has received widespread attention as a general probe of diffuse and translucent interstellar clouds and in particular as a surrogate for the H$_{2}$ column density determinations in such environments  \citep[e.g.,][and references therein]{federman1982measurements, sheffer2008}. CH has been observed in widely different wavelength regimes, from the radio at 9~cm \citep{rydbeck1973} over the sub-millimetre (sub-mm) and far-infrared (FIR) ranges, at 560~$\mu$m \citep{gerin2010} and 150~$\mu$m \citep{stacey1987,wiesemeyer2018unveiling} and the optical \citep[4300.3 \AA. e.g.,][]{Danks1984, sheffer2008} into the far-ultraviolet (FUV) regimes \citep[1369.13~\AA,][]{watson2001assignment}. In fact, the famous 4300.3 \AA\ CH transition was one of the first three molecular lines that were detected in the interstellar medium (ISM) \citep{dunham1937interstellar, swings1937}. While an abundance of $^{12}$CH data exists, very little is known about its rarer isotopologue ${}^{13}$CH. Unlike its parent molecule, which is ubiquitously distributed in the ISM, the only known astronomical identification of \ttch has been made in the solar spectrum by \citet{richter1967bestimmung}. As to the ISM, \citet{bottinelli2014ch} report the non-detection of the $N,J=1,1/2\rightarrow1,3/2$ and $N,J = 1,3/2 \rightarrow 2,5/2$ transitions of \ttch toward the well studied low-mass protostellar condensation IRAS 16293$-$2422. 

The \cratio ratio has been widely studied toward molecular clouds in the Milky Way \citep[e.g.,][]{henkel1982further, steimle1986lambda,wilson1994abundances, henkel1994interstellar} and, recently, also in the nuclear regions of nearby starburst galaxies \citep{tang2019}. It is an important diagnostic tool for probing Galactic chemical evolution or simply the nucleosynthesis history of the Galaxy. $^{12}$C is synthesised as the primary product of shell and core He-burning in intermediate- and high-mass stars via the triple-$\alpha$ reaction, while $^{13}$C is a secondary product of stellar nucleosynthesis and is produced over longer timescales. It is predominantly formed as a by-product of the carbon-nitrogen-oxygen (CNO) cycle in asymptotic giant branch (AGB) stars. Initiated by the proton capture of the $^{12}$C nucleus produced from an older stellar population, the CNO-cycle forms $^{13}$N which then decays via positron emission to form the $^{13}$C nucleus ($^{12}$C(p,$\gamma$)$^{13}$N($\beta^{+}$)$^{13}$C) \citep{pagel1997nucleosynthesis}. The $^{13}$C intermediate product is injected into the ISM via mass loss of the AGB stars, after the ashes of the helium burning shell of these objects have intermixed with their convective envelopes in the third ``dredge-up'' \citep{herwig2005evolution}. This establishes the \cratio isotopic ratio as a measure of the degree of astration present in the Galaxy. Chemical evolution models of the Galaxy \citep{tosi1982cno, prantzos1996evolution} have predicted the \cratio ratio to exhibit a positive gradient, increasing with Galactocentric distances and decreasing with time. 

The predictions of these models have been confirmed by observational measurements of the \cratio isotopic ratio carried out by studying the rotational transitions of molecules like H$_{2}$CO \citep{henkel1982further}, CO \citep{langer1990c}, and CN \citep{milam200512c} and of their respective ${}^{13}$C isotopologues, at cm and mm wavelengths\footnote{Where the lines from the isotopologues are well separated unlike in the optical regime (except for $^{12/13}$CH$^+$ see, \citet{ritchey2011interstellar}).}. While the average fit to the \cratio gradients derived independently using the three molecules mentioned above are in agreement within the quoted error bars, their individual trends display systematic variations amongst themselves. Reasons for these variations may be related to the observations of the different tracers and/or to isotope-selective chemical processes like gas-phase fractionation and selective photo-dissociation which do not impact every molecule in the same way. Chemical fractionation occurs as a result of ion-molecule exchange reactions that preferentially transfer and incorporate the heavier atomic isotope into molecules due to differences in zero-point energies between the different isotopes \cite[see, e.g.,][]{wilson1999isotopes, roueff2015}. Isotopic fractionation does not impact all molecules in the same way as its degree greatly depends upon the formation pathway of the molecule and the environment in which it is formed and exists. In molecules that are susceptible to fractionation, like CO, the chemical fractionation enriches the $^{13}$C isotope and lowers the \cratio ratio while it is increased by selective photo-dissociation of the rarer isotopic species containing $^{13}$C, due to its weaker self-shielding in regions with a large UV flux \citep{bally1982isotope}. Therefore, in such regions the $^{12}$CO/$^{13}$CO ratio might be higher than the underlying \cratio ratio. In denser regions where carbon exists mostly in the form of CO, $^{13}$C is locked up in CO and is depleted in other carbon-bearing molecules \citep{roueff2015}. Precise measurements of the \cratio isotopic ratio have also been made using observations of the CH$^{+} A-X$~(0, 0) transition near 4232~\AA\  \citep{ritchey2011interstellar} but their observations are limited to nearby stars (\textless 7~kpc) that are bright ($V$ \textless 10 mag) in the visible wavelength range. The corresponding sub-mm transitions of CH$^{+}$ are often optically thick with saturated absorption profiles and can only yield lower limits on the carbon isotopic ratio \citep{falgaronechp}. Hence, the high optical depths of many abundant $^{12}$C bearing molecules, and effects of saturation and self absorption pose a problem when using the intensities of such lines, which in turn skews estimates of the \cratio ratio. In an attempt to provide additional constraints on the \cratio ratio, we here propose the use of a new tracer -- CH.

 CH should be a good candidate for conducting isotopic ratio measures because it is an abundant species that is ubiquitously formed in the ISM and its spectral lines are predominantly optically thin. In particular, the hyperfine structure (hfs) components of the  $N,J=1,1/2\rightarrow2,3/2$ transitions of CH near $2\,$THz have recently been observed in generally unsaturated absorption, ideal for column density determinations \citep{wiesemeyer2018unveiling,jacob2019fingerprinting}. Models and simulations by \citet{rollig2013carbon} for photo-dissociation regions (PDRs) show that the fractionation present in CH is dominated by the fractionation of its parental species. CH originates from C$^{+}$ and is formed via the dissociative recombination of CH$_{3}^+$:
\begin{equation}
        \text{C}^+ \xrightarrow{\text{H}_{2}} \text{CH}_{2}^+  \xrightarrow{\text{H}_{2}} \text{CH}_{3}^{+} \xrightarrow{\text{e}^{-} } \text{CH} \, 
\end{equation}
Hence, the degree of fractionation in CH is closely coupled with that of C$^{+}$. At low visual extinctions, $A_\text{V}$, the isotopic fractionation ratio of C$^{+}$\footnote{\citet{Ossenkopf2013} have studied the fractionation of C$^{+}$ by comparing model predictions with high-resolution observations of the fine structure lines of [$^{12}$C{\small II}] and [$^{13}$C{\small II}] using HIFI/\textit{Herschel}. } is comparable to the elemental abundance ratio, however, it increases with $A_\text{V}$ at low gas temperatures typically by a factor of at most ${\sim}$2. This is a consequence of the enrichment of $^{13}$C in CO in these regions (favouring the forward reaction of the C$^{+}$--~CO ion-molecule exchange reaction \citep{langer1990c}), which results in the depletion of $^{13}$C$^{+}$ and all subsequent species like $^{13}$CH that are formed from it. In addition, the abundance of CH is often enhanced through hydrogen abstraction reactions of CH$^{+}$, which is formed endothermically ($\Delta E/k_\text{B}$ = 4620~K) \citep{hierl1997rate}, in shocks or the dissipation of turbulence \citep{godard2014chemical}, followed by the dissociative recombination of CH$_{2}^+$:
\begin{equation}
    \text{C}^+ \xrightarrow[\text{H}_{2}, \, h\nu]{\text{H}} \text{CH}^+ \xrightarrow[\text{H}_{2}, \, h\nu]{\text{H}}
    \text{CH}_{2}^{+} \xrightarrow{\text{e}^{-} } \text{CH} \, 
\end{equation}
Being formed at high effective temperatures via non-thermal processes, CH$^{+}$ is not affected by fractionation and is believed to reflect the ambient C-isotopic ratio in the diffuse ISM. Hence, in low $A_\text{V}$ diffuse clouds, where turbulence driven reactions are favoured, CH, like its parent CH$^{+}$, is not affected by fractionation. Therefore, the abundance ratio of the CH isotopologues does not deviate from the elemental ratio of the atomic carbon isotopes.  

As mentioned above, optical and UV absorption studies of CH require visually bright early type stars as background whose light is unhampered by interstellar extinction \citep{federman1982measurements, sheffer2008}. Owing to this, they can only probe the ISM in the solar neighbourhood out to a few kpc. In contrast, we observe supra-THz transitions of CH and \ttch in absorption against the bright dust emission of far away star-forming regions (SFRs) out to the Galactic centre (GC) and beyond. Thus our absorption spectra not only probe the molecular envelopes of these sources, but also the diffuse and translucent interstellar clouds along their lines of sight over a wide range of Galactocentric radii. Due to the Galaxy's differential rotation, the absorption covers wide local standard of rest (LSR) velocity ranges. As a caveat, we mention that given, first, the expected  ${}^{12}$CH/ ${}^{13}$CH ratios, which range from ${\approx}20$ in the GC to ${\approx}90$ at the Solar circle and even higher values beyond \citep[and references therein]{wilson1994abundances} and, second, that even ${}^{12}$CH lines are predominantly optically thin\footnote{Quite remarkably, this is actually generally true for radio, sub-mm, FIR and optical CH lines, except for the ground state rotational sub-mm/FIR transitions near 533/536, and 2006/2010 GHz, which reach optical depths exceeding unity in absorption from the envelopes of star forming regions; see, e.g., our Fig.\ref{fig:ch_spec_1}.} we expect a detection of ${}^{13}$CH absorption only toward the LSR velocities of the star forming regions studied, because these sources have far greater column densities than the intervening diffuse clouds along their lines of sight. In this paper, we present our search for \ttch in the ISM along the line-of-sight (LOS) toward five high-mass SFRs in the Milky Way, Sgr B2(M), G34.26+0.15, W49(N), W51E and W3(OH) and discuss its use as an unbiased tool for benchmarking the \cratio Galactic gradient.

\section{\texorpdfstring{${}^{13}$CH spectroscopy}{13CH spectroscopy}}
\begin{figure}
    \centering
    \includegraphics[width=0.48\textwidth]{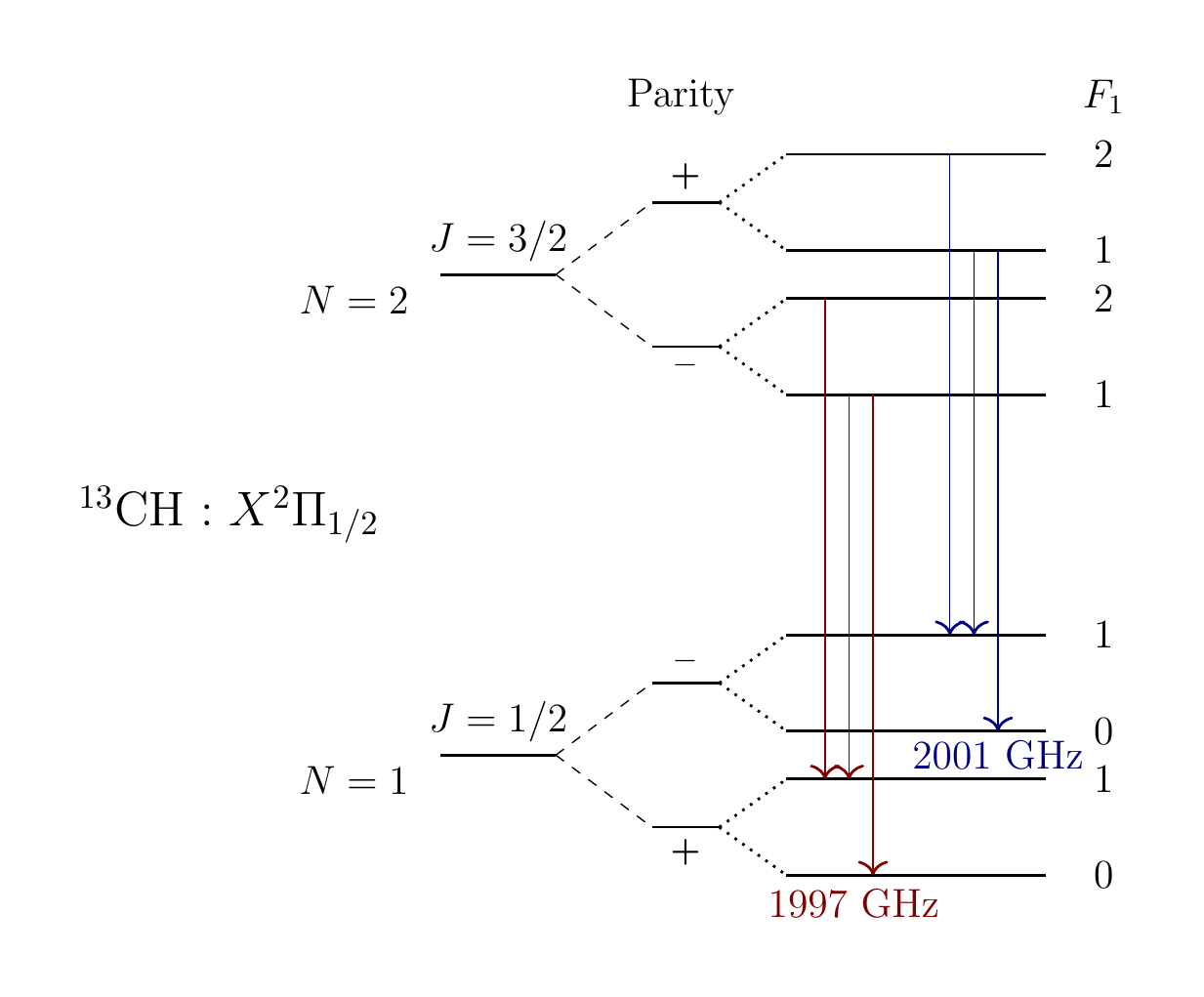}
    \caption{Energy-level diagram of the lower rotational levels of ${}^{13}$CH. The hyperfine transitions of \ttch that are presented in this study are highlighted using red and blue arrows. The proton hyperfine structure splittings are not included. Note that the level separations are not to scale.}
    \label{fig:energy_level}
\end{figure}

Similar to CH, ${}^{13}$CH conforms to Hund’s case b coupling but differs in the nuclear spin, $\boldsymbol{I}$, of the Carbon atom, $I$(${}^{12}$C)=0 and $I$(${}^{13}$C)=0.5. Due to the non-zero nuclear spin of the ${}^{13}$C isotope, the total angular momentum $\boldsymbol{J}$ first couples with $I_{1}$(${}^{13}$C) to generate $\boldsymbol{F_{1}}$ ($\boldsymbol{= J + I_{1}}$) which further couples with the nuclear spin of the H atom, $I_{2}$(H) = 0.5 to yield $\boldsymbol{F}$ ($\boldsymbol{= F_{1} + I_{2}}$). The energy level diagram of the ${}^{13}$CH $\Lambda$-doublet transitions that are discussed in this study are displayed in Fig.~\ref{fig:energy_level}. 
Considerable effort has been expended in the laboratory to measure the rotational spectrum of $^{13}$CH. The spectroscopic parameters of the rotational transitions of the \ttch radical between the $X^{2}\Pi$ ground state have been measured using the technique of laser magnetic resonance (LMR) at FIR wavelengths by \citet{davidson2004far}. The results of their experiments were combined with previously determined $\Lambda$-doublet intervals of the molecule by \citet{steimle1986lambda} to provide accurate predictions of the transition frequencies between the low-lying rotational levels and the ground state. The spectroscopic parameters of the observed $N,J = 2, 3/2 \rightarrow 1, 1/2$ transitions are tabulated in Table~\ref{tab:spec_params}. The Einstein A coefficients, $A_{\text{E}, ij}$, were computed from the line strengths, $S_{ij}$, using the relation:
\begin{equation}
    A_{\text{E}, ij} = \left( 16\pi^{3}\nu_{ij}^{3}/3\epsilon_{0}hc^{3}\right)\left( 2F_{i} + 1 \right)^{-1} S_{ij}\,\mu^2 \, 
    \label{eqn:einsteinA}
\end{equation}
for spontaneous emission from an energy level $i$ to $j$. Where $\nu_{ij}$ is the corresponding frequency, $F_{i}$ is the total hyperfine quantum number, of level $i$ and $\mu$ represents the electric dipole moment of ${}^{13}$CH, $\mu({}^{13}\text{CH})=$1.46~Debye \citep{pickett1998submillimeter}.

\begin{table}
   
            \caption{Spectroscopic parameters for the $N,J= 2, 3/2 \rightarrow 1, 1/2$ hyperfine structure transitions of \ttch. Taken from \citet{davidson2004far}.}
    
    \begin{tabular}{ccccc}

    \hline \hline
        \multicolumn{2}{c}{Transition} &  Frequency  & $S$ & $A_\text{E}\times10^{-3}$\\
          Parity & $F_{1}^{\prime},F^{\prime} \rightarrow F_{1}^{\prime\prime},F^{\prime\prime}$&  [GHz] &  & [s$^{-1}$] \\ \hline
         
           $+ \rightarrow -$ & $1,3/2\rightarrow 0,1/2$ & 1997.4232 & 0.3331 & 1.1458\\
                             & $1,3/2\rightarrow 1,3/2$ & 1997.4464 & 0.1666 & 0.5731   \\
                             & $2,5/2\rightarrow 1,3/2$ & 1997.4437 & 0.8327 & 1.9096\\ \hline
          $- \rightarrow +$  & $1,3/2\rightarrow 0,1/2$ & 2001.5672 & 0.3345 &  1.1577\\
                             & $1,3/2\rightarrow 1,3/2$ & 2001.2230 & 0.1673 &  0.5787\\
                             & $2,5/2\rightarrow 1,3/2$ & 2001.3673 & 0.8365 & 1.9296\\ \hline 
    \end{tabular}
    \label{tab:spec_params}
   \end{table}

\section{Observations}\label{sec:obs}
Using the upGREAT instrument\footnote{The German REceiver for Astronomy at Terahertz frequencies (GREAT) is a development by the MPI f\"{u}r Radioastronomie and the KOSMA/Universit\"{a}t zu K\"{o}ln, in cooperation with the DLR Institut f\"{u}r Optische Sensorsysteme.} \citep{risacher2016upgreat} on board SOFIA \citep{young2012early}, we observed the ${}^{2}\Pi_{1/2} \, N, J = 2, 3/2 \rightarrow 1, 1/2$ $\Lambda$-doublet transitions of \ttch over several flight series as a part of the observatory's cycle 7 campaign (under the open time project 07\_0148 supplemented by guaranteed time observations). In this pilot study we carried out observations toward five well known SFRs Sgr~B2(M), G34.26+0.15, W49(N), W51E, and W3(OH). Given that so far $^{13}$CH has never been detected in the ISM before, the primary source selection criterion was the existence of a strong sub-mm and FIR background continuum. Secondly, we selected sources that are almost evenly spaced in Galactocentric distance between the GC and the Solar circle in order to obtain quantitative constraints on the Galactic \cratio abundance ratio gradient. Observational properties of the different sources are summarised in Tab.~\ref{tab:source_properties}. The receiver configuration is comprised of the (7+7) pixel low frequency array (LFA) receiver module of upGREAT, in dual polarisation. Only data from the central pixels were used because the continuum targets are unresolved in our 13.5$^{\prime\prime}$ FWHM beam\footnote{The beam-size is scaled with frequency from the nominal 14.1$^{\prime\prime}$ beam width at the [C{\tiny II}] 158~$\mu$m line.}. The spectra were taken in the double-beam switch mode, chopping at a frequency of 2.5~Hz with a chop throw between 210$^{\prime\prime}$ and 240$^{\prime\prime}$ and a chop angle of 90$^{\circ}$ (counter-clockwise against North), to account for both atmospheric fluctuations, as well as fluctuations that may arise from the instrument. The receiver was connected to an evolved version of the MPIfR Fast Fourier Transform Spectrometer described by \citet{klein2012}. This backend provides a $4\,$GHz bandwidth per pixel and a velocity resolution of $0.036$~km~s$^{-1}$ (${\sim} 244.1$~kHz) over 16384 channels. In this detection experiment, the double-sideband (DSB) receiver was tuned to 1997.4437~GHz (the strongest hfs transition of the 1997~GHz $\Lambda$-doublet) in the lower sideband. Three different intermediate-frequency (IF) settings were used in our observations in order to disentangle any contamination present in the bandpass, toward all sources except W3(OH) toward which we used only a single IF setting at 1.45~GHz in our pilot search. For G34.26+0.15, W49(N) and W51E the IF was tuned to 1.2, 1.4 and 1.6~GHz while for Sgr B2(M) the IF was tuned to 1.4, 1.6, and 1.8~GHz. 
The \ttch spectra obtained along the LOSs to Sgr B2(M), G34.26+0.15, W51E and W49(N) taken with each of the three IF settings are displayed in Appendix.~\ref{appendix:dsb_ceonvolution}.
\begin{table*}
\centering 
    \caption{Continuum source parameters.}
    \begin{tabular}{l rr   lrr rrr}
    \hline\hline
    Source &  \multicolumn{1}{c}{$\alpha$(J2000)} & \multicolumn{1}{c}{$\delta$(J2000)} &
    \multicolumn{1}{c}{$\upsilon_{\text{LSR}}$} & \multicolumn{1}{c}{$T_{\text{c}}$}  &  $R_{\text{GC}}$ & \multicolumn{1}{c}{$D$}  & \multicolumn{1}{c}{Flight Id} & $t_{\text{obs}}$\tablefootmark{a}\\
    & \multicolumn{1}{c}{[h:m:s]} & \multicolumn{1}{c}{[$^{\circ}:^{\prime}:^{\prime\prime}$]} 
    &  [kms$^{-1}$] & [K] & [kpc] & [kpc] &  & [mins] \\
    \hline
         Sgr B2(M) & 17:47:20.49 & -28:23:06.00 
          & 65.2(0.6) & 14.6 & 0.1 & 8.3  & 2019/06/10(F579) &  51\\
         & & & & & & &  2019/06/11(F580) & 101\\
         
         W51E & 19:23:43.90 & +14:30:30.50
         & 63.0(1.6) & 9.4 & 6.3 &  5.4  &
         2019/12/11(F646) &  90\\
         G34.26+0.15 & 18:53:18.49 &  +01:14:58.70 
         &  58.0(0.7) & 8.6 & 7.0 & 1.6  &  2020/03/07(F669) & 126\\
         & & & & & &  &  2020/03/11(F671) &   85 \\
         W49(N) &  19:10:13.20 & +09:06:11.88 
         & 11.8(0.4) & 13.8&  7.8 & 11.4 &   2020/03/05(F667) & 80\\
         & & & & & &  &  2020/03/10(F670) &  55\\
         & & & & & & &   2020/03/12(F673) &  135\\
         & & & & & & &    2020/03/13(F674) & 57 \\
         W3(OH) & 02:27:04.10 & +61:52:22.00 &-46.0(0.2) & 5.3 & 10.0 & 2.0 & 2018/11/21(F529) &  117\\
         & & & & & &   &  2018/12/04(F533) & 53\\
         \hline
    \end{tabular}
    \tablefoot{Columns are, left to right, source designation, equatorial coordinates, LSR velocity, signal band continuum brightness temperature derived by means of a DSB calibration, Galactocentric distance, heliocentric distance, flight id and flight leg duration. \tablefoottext{a}{The observing leg time ($t_{\text{obs}}$) refers to the total duration of time for which the source was observed. This includes the total (on+off) observing time as well as the overheads, which in total is typically a factor of two larger than the (on+off) time.}}
     \tablebib{References for the heliocentric distances: Sgr B2(M):~\citet{reid2014trigonometric};
     W51E:~\citet{sato2010};
     G34.26+0.15:~\citet{zhang2009trigonometric};
     W49(N):~\citet{zhang2013parallaxes};
     W3(OH):~\citet{xu2006distance};}
    \label{tab:source_properties}
    \end{table*}
Using a forward efficiency of 0.97 the spectra were further calibrated using the KALIBRATE program \citep{guan2012great}. Fluctuations of the continuum level can either be due to telescope tracking problems, or due to gain drifts in the mixers. From a comparison between the continuum fluxes in the high-frequency array, operated in parallel at 4.7~THz and offering a 6$^{\prime\prime}$ FWHM beam, we can rule out the former (which even if they occurred would not have an impact on the line-to-continuum ratio). Gain drifts that are faster than the calibration rate, however, affect the measurement of the atmospheric total power and therefore the applied transmission correction. Since they potentially distort the line-to-continuum ratio, a new calibration strategy was applied to analyse the gain fluctuations of the GREAT receiver, thanks to the larger amount of data now available. In a first step, the most stable off-centre pixels were identified by monitoring the line areas of a simultaneously observed narrow telluric ozone line, which is largely insensitive to baseline uncertainties. Since the atmospheric emission arises in the near-field of all pixels, they must, therefore, all see the same ozone line flux. From this correlation analysis, only those pixel pairs whose line flux ratios were persistently close to unity were retained to determine the atmospheric transmission correction. The quality of the correlation analysis was ensured by forming closure products of the gain ratios, including those of the central pixel used (which for our case deviates from unity by at most 1.6\%). In order to eliminate spectra in the central pixel affected by gain drifts, only data with closure products deviating from unity by at most 0.4\% were retained to determine the continuum level. The DSB-continuum level is then determined by accounting for contributions from both the signal and image bands, which is added back to the spectra to obtain the correct line-to-continuum ratio. This calibration technique was applied not only to the \ttch data presented in this work but also to the previously published $^{12}$CH data \citep{wiesemeyer2018unveiling,jacob2019fingerprinting}, which thus were re-calibrated for use in our analysis, to avoid any inconsistencies. For the $^{12}$CH spectra, we note that the continuum levels derived using the closure products are compatible with those derived using the standard calibration methods, except for W49(N). Using the closure product analysis we derive a continuum level of 14 K for the $^{12}$CH spectrum toward W49(N) which agrees with the continuum level cited by \citet{wiesemeyer2018unveiling} within a 20\% uncertainty. It is to be noted that, in the analysis that follows, we adopt a value of 14~K as the continuum level of the $^{12}$CH spectrum toward W49(N).

\begin{figure*}
\centering
\includegraphics[width=0.9\textwidth]{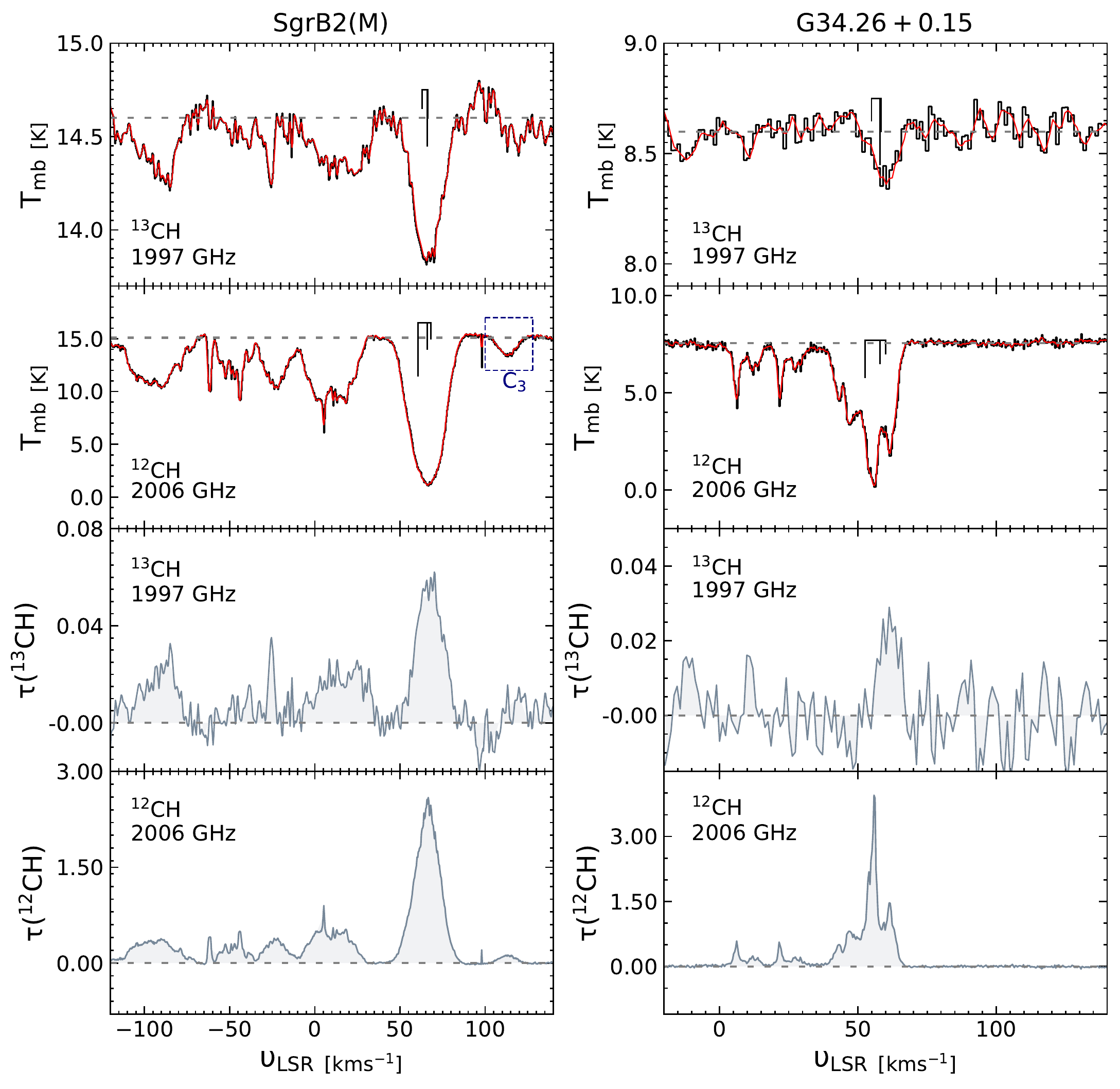}\\
\caption{ From top to bottom: $N,J = 2,3/2 \rightarrow 1,1/2$ transition spectra of \ttch and CH near 1997~GHz and 2006~GHz, in black with their corresponding Wiener filter fits to the spectra overlaid in red toward Sgr~B2(M) (left) and G34.26+0.15 (right). Followed by their Wiener filter deconvolved spectra displayed in grey in optical depth scales. Note: The $^{12}$CH spectrum toward Sgr~B2(M) is contaminated by C$_{3}$ absorption at 2004.833~GHz in the image band as indicated in blue.}
\label{fig:ch_spec_1}
\end{figure*}

\begin{figure*}
\centering
\includegraphics[width=0.9\textwidth]{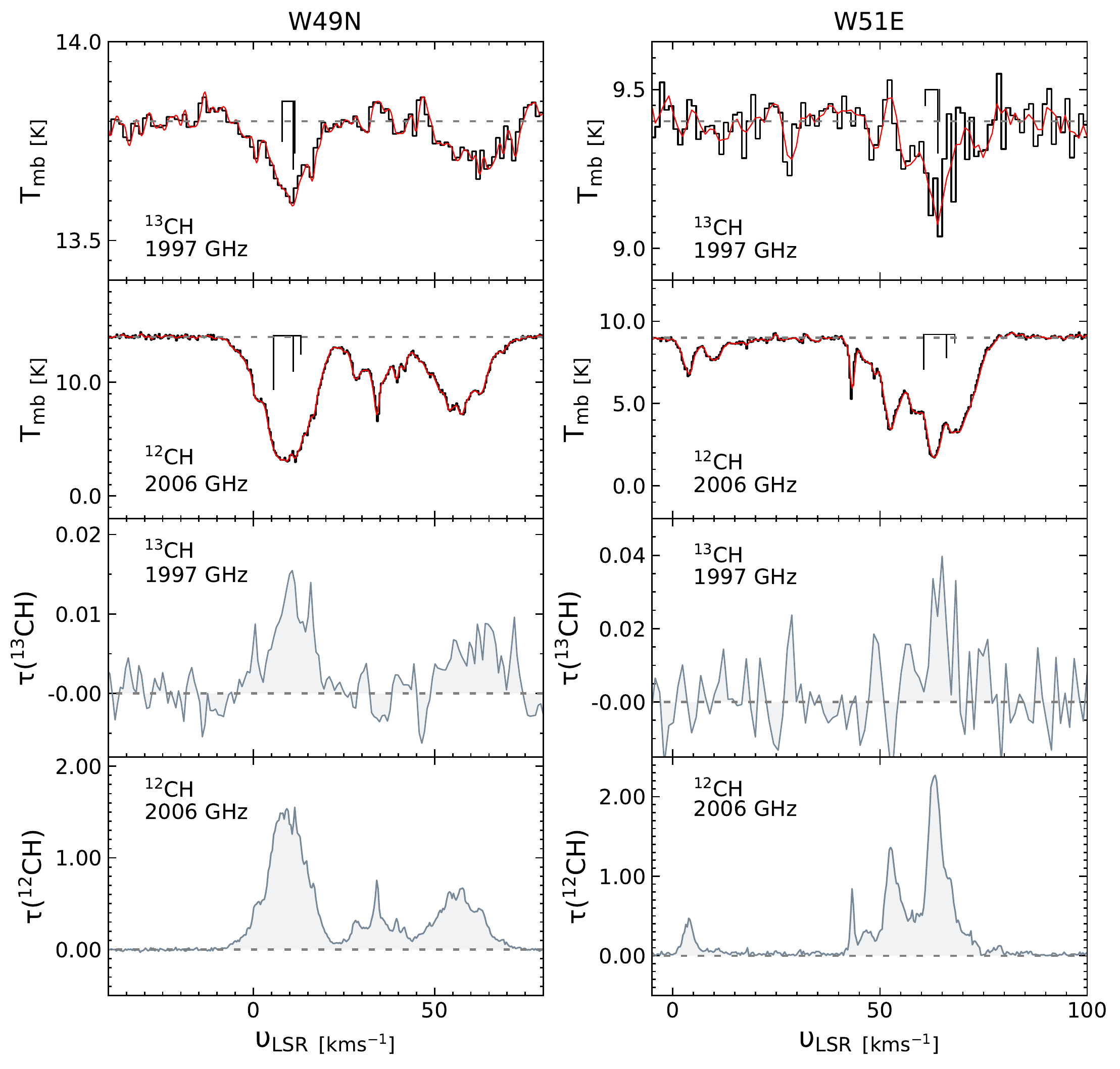}\\
\caption{ Same as Fig.~\ref{fig:ch_spec_1} but toward W49(N) (left) and W51E (right).}
\label{fig:ch_spec_2}
\end{figure*}

Subsequently, the fully calibrated spectra obtained from both polarisations of the central pixel were converted to main-beam brightness temperature scale (using a main beam efficiency of 0.66) and analysed using the GILDAS-CLASS software\footnote{Software package developed by IRAM, see \url{https://www.iram.fr/IRAMFR/GILDAS/} for more information regarding GILDAS packages.}. The spectra were box-smoothed to ${\sim}1.1$~km~s$^{-1}$ wide velocity bins and the spectral baselines were corrected for by removing up to a second order polynomial. The resultant spectra obtained after carrying out a simple sideband deconvolution (discussed in Appendix~\ref{appendix:dsb_ceonvolution}) are displayed in Figures~\ref{fig:ch_spec_1} and \ref{fig:ch_spec_2}. On average we achieved a noise level of ${\sim}$68~mK at a velocity resolution of 1.1~km~s$^{-1}$ after the sideband deconvolution. In the following paragraphs we briefly describe the LOS properties of the observed \ttch spectra and compare them with previously obtained CH spectra.

Sgr B2(M): The unwavering nature of the absorption feature at $\upsilon {\sim} 64~$km~s$^{-1}$ observed toward the envelope of Sgr B2(M) in each of the different IF frequency settings and the detection of its image band $\Lambda$-doublet counterpart near 2001~GHz with the absorption features displaying their expected relative intensities, solidifies our detection of \ttch near 1997.44~GHz at this velocity. We compare the \ttch spectrum with that of CH towards Sgr B2(M) published in \citet{jacob2019fingerprinting}. The CH spectrum has a true continuum of 15~K and is contaminated by C$_{3}$ absorption at 2004.833~GHz arising from the image band as indicated by the blue box in Fig.~\ref{fig:ch_spec_1}. The similarities between the CH and \ttch spectra suggest the presence of weak blue-shifted sight-line absorption features in addition to the deep absorption seen near 64~km~s$^{-1}$ corresponding to the systemic LSR velocity of the molecular cloud. The shift of the atmospheric ozone feature at 2002.347~GHz in the image band toward the signal band features, by 60~km~s$^{-1}$ with each 0.2~GHz IF offset, leads to uncertainties while fitting polynomial baselines particularly for the weaker features. Hence, the true nature of the sight-line absorption features remains uncertain because of calibration and baseline uncertainties for the broad blended features. We will follow this up in future observations. 

G34.26+0.15: In order to avoid the blending of the atmospheric ozone feature from the image band with the signal band features, we used an IF setting of 1.2~GHz instead of 1.8~GHz. Similar to the \ttch spectrum observed toward Sgr B2(M), we see absorption at the systemic velocity of the G34.26+0.15 molecular cloud at 58~km~s$^{-1}$, in all three IF settings. However, we do not clearly detect \ttch in foreground gas at velocities between 0 and 50~km~s$^{-1}$.

W49(N): We used a setup similar to that used for our observations toward G34.26+0.15 for those toward W49(N). In addition to \ttch absorption at the velocities corresponding to that of the molecular cloud, we see a weaker absorption feature close to 65~km~s$^{-1}$, associated with the far-side crossing of the Sagittarius spiral arm. However, we do not include this feature in the analysis that follows because of its low signal-to-noise level.

W51E: By choosing the 1.4~GHz IF tuning as the nominal setting for carrying out the DSB deconvolution, we observe \ttch absorption at the source intrinsic velocity of W51E near ${\sim}$62~km~s$^{-1}$. Comparing this with the corresponding $^{12}$CH spectrum which closely resembles that of the ${}^{2}\Pi_{3/2}, J = 3/2 \rightarrow 5/2$ SH line at 1383.2~GHz, between 43--80~km~s$^{-1}$ \citep{neufeldsulphur}, it is not clear whether \ttch shows a second absorption component at 53~km~s$^{-1}$ at the given noise level. The weaker and narrower absorption seen at 26~km~s$^{-1}$ is merely a remnant feature from the image band.

Toward W3(OH) no \ttch lines are detected above the noise (77~mK at a spectral resolution of 1.1~km~s$^{-1}$) in the spectrum at a continuum level of 5.3~K. The baseline-subtracted CH and \ttch spectra toward W3(OH) are displayed in Appendix~\ref{fig:w3oh_spec}. The corresponding CH spectrum is taken from \citet{wiesemeyer2018unveiling} and has a continuum level at 5.4~K, prior to baseline removal.

Additionally we have checked whether the $^{12}$C/$^{13}$C ratios derived by us are consistent with values derived using archival HIFI/\textit{Herschel} data of the $N, J =
1,3/2 \rightarrow 1, 1/2$ transitions of CH and $^{13}$CH near 532~GHz. While the 532~GHz CH line displays a deep absorption feature at the systemic velocity of the Sgr B2(M) molecular cloud with a peak temperature of 1.64~K \citep{qin2010},
the corresponding $^{13}$CH transition is not detected above a 3$\sigma$ noise level of 0.33~K (at a spectral resolution of 0.5~km~s$^{-1}$) and yields a lower limit of 5 on the $^{12}$C/$^{13}$C isotopic ratio. Similarly, we have analysed the 532~GHz CH spectra taken toward G34.26+0.15 discussed in \citet{Godard2012}, W49(N), and W51E presented in \citet{gerin2010} and W3(OH) from the Herschel archives\footnote{See, \url{http://archives.esac.esa.int/hsa/whsa/}} 
that also cover the corresponding \ttch line frequencies. We find no signatures of $^{13}$CH and, from the 3$\sigma$ noise levels, are only able to derive lower limits to the $^{12}$C/$^{13}$C ratio of 13, 32, 17, 68 for the above sources. We mention that, in contrast to the 2~THz lines, in the 532 GHz line, the molecular cores associated with these massive SFRs show complex line profiles with both emission and absorption components. In addition, the intensities and profiles of the \ttch lines are quite uncertain as their HIFI spectra are affected by a standing wave and an unknown (and not yet assessed) level of potential line contamination. Thus the formal lower limits for the  $^{12}$C/$^{13}$C ratio derived from archival HIFI/\textit{Herschel} data quoted above should be regarded with some caution.

\section{Analysis and Discussion}

High-resolution absorption line spectroscopy provides a powerful and straightforward to use tool for measuring column densities. The optical depth, $\tau$, for a single absorption component can be calculated from the line to continuum ratio using
\begin{equation}
    \tau = -\text{ln}\left( T_\text{l}/T_\text{c}\right) \, ,
    \label{eqn:optical_depth}
\end{equation}
where $T_\text{l}$ and $T_\text{c}$ represent the observed brightness temperatures of the line (prior to continuum subtraction) and the continuum, respectively. We have determined the optical depth profile, i.e., $\tau$ vs. $\upsilon_{\rm LSR}$, using the Wiener filter fitting technique as described in \citet{jacob2019fingerprinting}. This fitting procedure first, fits the observed spectral profile by minimising the mean square error between the model and observations and then deconvolves the hyperfine structure from the observed spectrum using the hfs components' relative weights. Other than the observed spectrum and the spectroscopic parameters of the line to be fit, the only other input parameter required by the Wiener filter technique is the spectral noise, which is assumed to be independent of the observed signal. The resulting deconvolved optical depth signature, $\tau_{\text{decon}}$, can then be converted into column density values per velocity channel, $i$ using
\begin{equation}
    \left( \frac{\text{d}N}{\text{d}\upsilon} \right)_{i} = 
\frac{8\pi\nu_{i}^3}{c^3}\frac{Q_{\text{rot}}}{g_{\text{u}}A_\text{E}}\text{exp}\left(\frac{E_\text{u}}{T_\text{ex}}\right)\left[ \text{exp}\left(\frac{h\nu_{i}}{k_\text{B}T_\text{ex}}\right)-1\right]^{-1} \left(\tau_{\text{decon}}\right)_{i} \, ,
\label{eqn:column_density}
\end{equation}
where the spectroscopic parameters $g_\text{u}$ (the upper level degeneracy), $E_\text{u}$ (the upper level energy) and $A_\text{E}$ (the Einstein A coefficient) remain constant for a given hyperfine transition. The partition function, $Q$ is itself a function of the rotation temperature, $T_\text{rot}$, which would be equal to the excitation temperature, $T_\text{ex}$, under conditions of local thermodynamic equilibrium (LTE). Typically the rotational transitions of hydrides require large critical densities to be observable in emission. Since collisional rate coefficients are not presently available for $^{13}$CH, we assume the critical density of this \ttch line to be identical to the critical density of the corresponding CH transition, assuming a two-level system. Using hfs resolved collisional rate coefficients computed by \citet{dagdigian2018hyperfine}, we find the critical densities to be ${\sim} 2\times10^{9}~\text{cm}^{-3}$ at 50--100~K. Within the diffuse and translucent interstellar clouds along the LOS ($n = 10-500~\text{cm}^{-3}$) and the envelope of Sgr B2(M) (ranging from $n {\sim} 10^3$--$10^5~\text{cm}^{-3}$; \citep{Schmiedeke2016}) the gas densities are much lower than the critical density, $n_\text{crit}$, of these transitions making them sub-thermally excited. Hence we can assume the excitation temperature of such sub-thermally excited lines to be small and lower than the gas kinetic temperature. In our analysis we assume $T_{\text{rot}}$ to be equal to the temperature of the cosmic microwave background (CMB) radiation, $T_\text{CMB}$ = 2.728~K \citep{neill2014herschel}. Given this low rotation temperature and the inefficacy of purely collisional excitation, almost all of the CH molecules will reside in the molecule's ground state level. However, in the SFRs, radiative excitation by FIR dust radiation can affect the level populations (see, \citet{Neufeld1997} for the case of HF). Meaningful modelling of this is beyond the scope of the present study. In any case, given that the $^{12}$CH and $^{13}$CH lines under consideration here have, respectively, moderate and low optical depths, radiative excitation is expected to affect the observed transitions from both isotopologues in a similar way, leaving the $^{12}$CH/$^{13}$CH intensity ratio unaltered.

Given the knowledge of the frequency separation of the hfs transitions of $^{13}$CH,  we first deconvolve the hyperfine structure from the observed spectra using the Wiener filter formalism (briefly discussed above) and then derive column densities by adopting a $T_\text{ex}$ value equal to $T_\text{CMB}$ and integrating over the deconvolved velocity range of the line. We derive \ttch column densities between ${\sim}2\times10^{12}$ and $4.6\times10^{13}~\text{cm}^{-2}$ toward the different sources. Given that the optical depths are computed directly from the line-to-continuum ratio, the uncertainties in the true continuum level give rise to systemic errors in the derived column densities. We assume a 10$\%$ error in the continuum level calibration based on the instrumental performance \citep{kester2017derivation} and sideband dependence of the atmospheric transmission. The subsequently derived errors in the column densities (per velocity interval) are computed following the description presented in \citet{jacob2019fingerprinting} and scale with the deconvolved optical depths by a constant term comprised of the spectroscopic parameters that govern the transition and the excitation temperature. We compare the \ttch column densities derived here, over velocity intervals associated with the different molecular clouds, with those of their corresponding $N,J = 2, 3/2 \rightarrow 1, 1/2$ hyperfine hfs transitions of CH near 2007~GHz discussed in \citet{wiesemeyer2018unveiling} and \citet{jacob2019fingerprinting}. In Tab.~\ref{tab:spectra_params}, we present the derived ${}^{12}$CH and ${}^{13}$CH column densities, using the same excitation conditions by adapting Eq.~\ref{eqn:column_density}, as well as the resulting \cratio isotopic ratio. For W3(OH), toward which the \ttch line remains undetected, we derive a 2$\sigma$ lower limit on the ${}^{12}$CH/${}^{13}$CH abundance ratio of $>$ 58 over the velocity interval between (-55 -- -38)~km~s$^{-1}$.

\begin{table*}
    \centering
    \caption{Synopsis of the derived column densities and \cratio isotopic ratios}.    \begin{tabular}{lrrrrl}
         \hline \hline
         Source &  $\upsilon_{\text{min}} - \upsilon_{\text{max}}$ &$N(^{13}\text{CH})$ & $N(^{12}\text{CH})$ & \multicolumn{2}{c}{$^{12}$C/$^{13}$C}
          \\ 
         &  [km~s$^{-1}$] &  [$10^{13}~$cm$^{-2}$] & [$10^{13}~$cm$^{-2}$] & CH & CN\tablefootmark{a}  \\ \hline
         Sgr B2(M)  & 40 -- 90 & 4.61$^{+0.70}_{-0.71}$ &  73.02$^{+0.60}_{-0.68}$ & 15.8${^{+02.4}_{-02.4}}$ & \textgreater 12\\
         W51E & 45 -- 75& 1{\ .77}$^{+0.20}_{-0.26}$&  66.36$^{+3.30}_{-4.21}$ &  37.5$^{+04.6}_{-06.1}$& 35$\pm$12 \\
         G34.26+0.15 & 35 -- 70 & 0.20$^{+0.01}_{-0.02}$ &  9.38$^{+0.78}_{-0.93}$ &  47.0$^{+04.5}_{-06.6}$& 28$\pm$4\\
         W49(N) & -5 -- 25 & 0.68$^{+0.11}_{-0.09}$&  $ 22.87^{+3.56}_{-3.72}$ & $ 33.6^{+07.5}_{-07.0}$ &  44$\pm$22\\
         W3(OH) & -55 -- -38 & \textless0.11\tablefootmark{b}$\quad$& $ 6.58^{+0.72}_{-1.35}$& \textgreater$58\qquad \,\,$ & 63$\pm$16 \\
         \hline
    \end{tabular}
    \tablefoot{\tablefoottext{a}{Values taken from \citet{milam200512c} and references therein.}\tablefoottext{b}{The column density for the non-detection was derived using the 2$\sigma$ rms of level. }}
    \label{tab:spectra_params}
    
\end{table*}

The \cratio isotopic abundance ratio has been determined towards the GC region using a wide variety of molecules ranging from simple species like CO, CN, H$_{2}$CO and HCO$^{+}$ \citep{henkel1982further, langer1990c, savage2002galactic} to more complex ones containing more than six atoms like CH$_3$CH$_2$CN, CH$_{3}$CCH, CH$_{2}$CHCN and NH$_{2}$CHO \citep{belloche2016, halfen201712c} to name a few. The relatively lower values of the \cratio isotopic ratio towards the GC in comparison to the inner Galaxy, the local ISM and the solar system bears evidence to its advanced state of chemical evolution and reflects on its unique nucleosynthesis history. Moreover, the value of the GC strongly pivots the derived Galactic gradient. Therefore, as discussed by \citet{halfen201712c}, it is essential to obtain more accurate measurements of the \cratio ratio towards the GC region since several of the derived molecular isotopic ratios can be hindered by effects of optical depth and saturation, chemical fractionation and selective photo-dissociation. The limits on the \cratio isotopic ratio derived using CH are consistent with those derived by \citet{savage2002galactic} using CN for our sample of sources except for G34.26+0.15. The latter difference could be due to difficulties in deriving the ${}^{12}$CN optical depth (meaning, an underestimation of the column density) because, as the authors of the above study state, the observed relative line intensity ratios of the hfs components does not follow LTE. The comparable isotopic compositions of the two molecular species stems from their inter-linked formation routes\footnote{As the neutral-neutral reaction between CH and N forms one of the major formation pathways for the production of CN (CH + N $\rightarrow$ CN + H), with a forward reaction rate of ${\sim}6.7\times10^{-11}~\text{cm}^{3}~\text{s}^{-1}$ for a temperature of 50~K.} and suggests that CH similar to CN shows negligible amounts of fractionation. Further, \citet{halfen201712c} estimated \cratio isotopic ratios between 19 and 33 using several different complex organic species and an average value of ${\sim}24\pm9$ towards the 64~km~s$^{-1}$ component of the GC source Sgr B2(N). Within the errors, the average \cratio isotopic ratio derived by these authors is also consistent with the values derived from CH and CN. This presents solid evidence that the $^{13}$C isotopic enrichment in more complex molecules must arise from progenitor molecules like CH and CN since the $^{13}$C substitution of complex species via simple ion-molecule exchange reactions is not as straightforward as that of simple molecules.

In general, almost all the PDR models studied by \citet{rollig2013carbon} with varying physical parameters display a $^{12}$CH/$^{13}$CH fractionation ratio that is enhanced at higher values of $A_\text{v}$ ($\geq 1$). The degree of fractionation is coupled with the FUV flux present in the models, the weaker the FUV flux, the greater is the fractionation ratio. The \cratio isotopic ratios we derive from CH show no indication of an enhanced value in comparison to those derived from for e.g., CN  as shown in Tab.~\ref{tab:spectra_params} and therefore no signature of
fractionation. This is because toward the SFRs toward which we detect both $^{12}$CH and $^{13}$CH, the absorption from both isotopologues primarily traces these regions' \textit{extended} envelopes, which are exposed to a significant UV field and whose densities have been estimated to be of the order of $10^3$~cm$^{-3}$ on parsec scales for Sgr~B2(M) and other regions \citep{Schmiedeke2016,wyrowski2016}, a value typical for a translucent molecular cloud. The gas-phase carbon reservoir in such regions is predominantly in either its atomic or ionised form  and not locked up in CO, which means that there is enough $^{13}$C and $^{13}$C$^+$ present for ion-molecule exchange reactions to form $^{13}$C-substituted CH.

\begin{figure}
    \centering
    \includegraphics[width=0.45\textwidth]{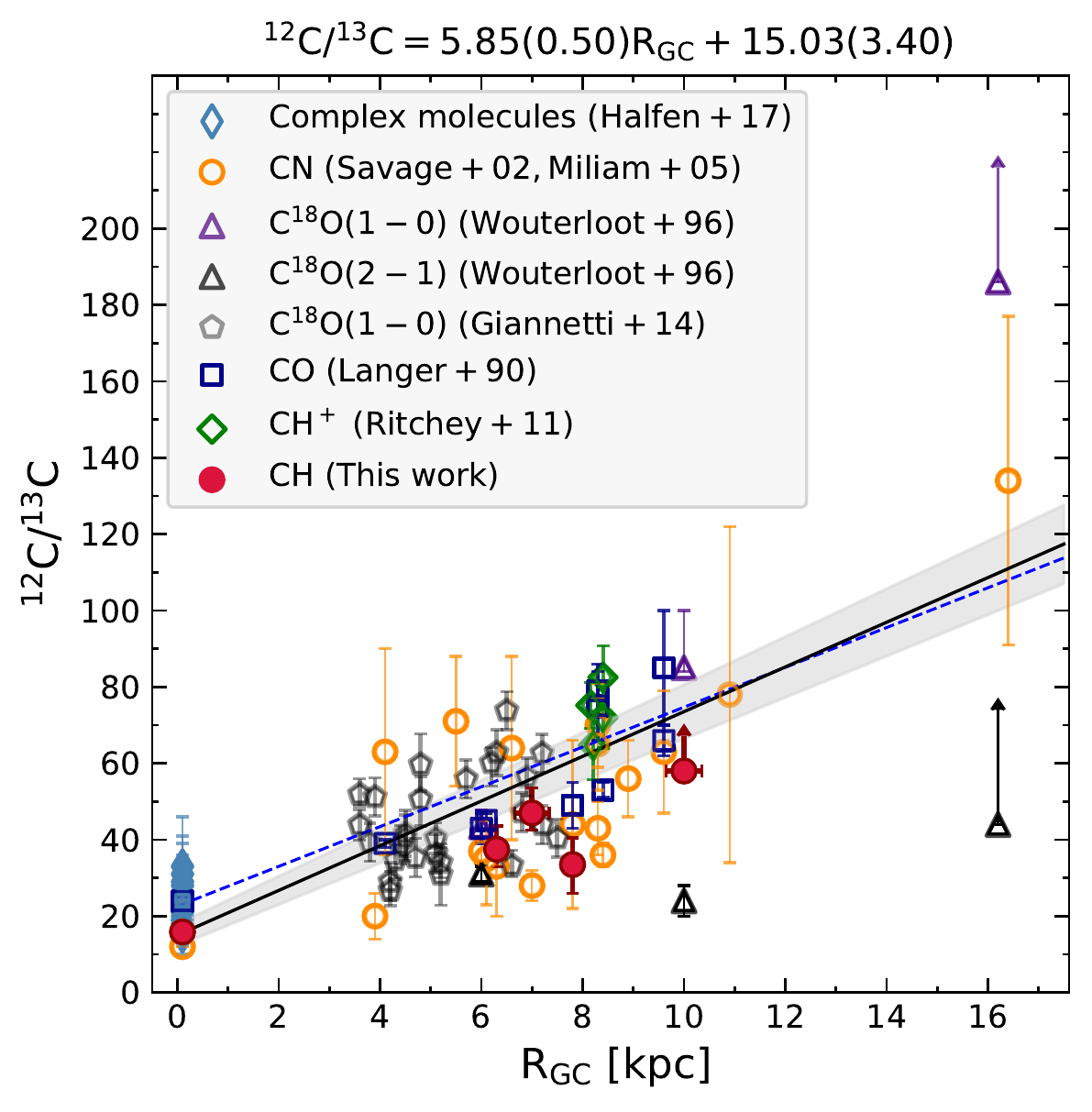}
    \caption{Plot of $^{12}$C/$^{13}$C isotope ratios as a function of Galactocentric distance, R$_\text{GC}$~(kpc). The filled red circles represent the \cratio ratio obtained using CH (this paper) while the unfilled black and purple triangles, grey pentagons, yellow circles, dark blue squares and green, and light blue diamonds are those obtained using isotopologues of C${}^{18}$O ( $J =1-0$ and $J=2-1$ transitions) \citep{wouterloot1996iras, Giannetti2014}, CN \citep{savage2002galactic,milam200512c}, CO \citep{langer1990c}, CH$^{+}$ \citep{ritchey2011interstellar}, and complex organic molecules \citep{halfen201712c}, respectively. The black solid line represents
the weighted fit to the data with the grey shaded region demarcating the 1$\sigma$ interval of this fit. For comparison the fit obtained by \citet{halfen201712c} is displayed by the dashed blue line.}
    \label{fig:12c13c}
\end{figure}

The previously determined \cratio isotopic ratios suffer from large error bars that may either be due to opacity effects in the main isotopologue or other systematic effects. Therefore it is not clear whether the large dispersion in values between Galactocentric radii of 4 to 8~kpc, corresponding to regions with the most molecular mass content in the Milky Way (apart from the GC region), are due to actual cloud-cloud variations. If the spread is indeed due to opacity effects then the ground state rotational transitions of CH studied in this work which are free from such effects, should be well-suited to quantitatively constrain the \cratio ratio. By combining the \cratio ratio values derived using CH with those derived by \citet{langer1990c}, \citet{wouterloot1996iras}\footnote{The authors of the cited article note that the discrepancy between the higher ratio derived form the $J = 1-0$ lines of $^{12}$C$^{18}$O and $^{13}$C$^{18}$O to the lower value from the $J=2-1$ lines ``is not yet explained, but may be due to the emission of the two transitions originating in different parts of the cloud with different excitation conditions.''}, \citet{milam200512c}, \citet{Giannetti2014}, \citet{ritchey2011interstellar}, and \citet{halfen201712c} and carrying out a weighted least squares fit, we derive a revised \cratio Galactic gradient of $\text{\cratio} =  5.85(0.50)~R_{\text{GC}} + 15.03(3.40)$\footnote{The values in parentheses represent 1$\sigma$ uncertainties.} displayed in Fig.~\ref{fig:12c13c}. The addition of our CH data points plus those from \citet{wouterloot1996iras}, \citet{ritchey2011interstellar}, and \citet{Giannetti2014} results in values for the best fit slope and intercept that are, within the combined uncertainties,
consistent with the values derived by \citet{halfen201712c}. The small uncertainties of the CH data result in somewhat smaller formal uncertainties of the fitted values.

\section{Conclusions}
In this paper we report the first detections of \ttch in the ISM, namely towards the Sgr B2(M), G34.26+0.15, W49(N), and W51E massive SFRs in the Milky Way. Hyperfine structure transitions connecting sub-levels of the \ttch $N, J= 2, 1/2$ and $1,1/2$ $\Lambda$-doublet states with frequencies near 1997~GHz were observed in absorption using GREAT/SOFIA, which provides an avenue to observe frequency bands for which spectrally resolved observations were previously not possible with HIFI/\textit{Herschel} or earlier missions. The detection of \ttch along with observations of the main isotopologue CH towards the same sources opens a new, independent way for determinations of the \cratio isotopic abundance ratio across the Galaxy. We derive \cratio isotopic ratios for those of our target sources with a \ttch detection and a lower limit for W3(OH) toward which we did not detect this isotopologue. Our values are in agreement with previous determinations made using varied chemical species and in particular CN. Our observations do not hint at a possible enhancement in the \cratio ratio derived from CH as it traces the more diffuse and translucent regions of the ISM in which CO is not the main reservoir of carbon. Furthermore, as its abundance peaks in regions of high UV radiation, CH is relatively unaffected by selective photo-dissociation and optical depth effects like saturation. Hence, measurements of the \cratio isotopic ratio based on the fundamental rotational lines of CH potentially reflect the actual \cratio ratio in the gas. 

In addition, knowledge of the ${}^{13}$C substitution in CH will improve our understanding of interstellar chemistry because direct substitution of ${}^{13}$C in more complex species is currently poorly understood and their observed $^{12/13}$C isotopologue ratios are speculated to have their origin in simpler precursors like CH. Investigating a larger sample of SFRs at different Galactocentric radii for \ttch will allow for a better constraint on the average \cratio abundance ratio value in the ISM and on the Galactic \cratio gradient, facilitating accurate Galactic chemical evolution models. However, the requirement of very high signal-to-noise ratios at high spectral resolutions that are required to detect the weak \ttch absorption lines greatly restricts the selection of sources to those with strong continuum backgrounds for carrying out future, follow-up studies. Further, we hope that our study will encourage coordinated laboratory efforts resulting in refinements of the spectroscopic parameters of $^{13}$CH, for which, e.g., several of the constants describing the fine structure of its transitions have not yet been well established \citep{halfen2008}.. 

\begin{acknowledgements}
We would like to thank David Neufeld and the anonymous referee for their valuable comments which helped to improve the clarity of this paper. SOFIA Science Mission Operations is jointly operated by the Universities Space Research Association, Inc., under NASA contract NAS2-97001, and the Deutsches SOFIA Institut under DLR contract 50 OK 0901 and 50 OK 1301 to the University of Stuttgart. upGREAT is financed by resources from the participating institutes, and by the
Deutsche Forschungsgemeinschaft (DFG) within the grant for the Collaborative Research Center 956, as well as by the Federal Ministry of Economics and Energy (BMWI) via the German Space Agency (DLR) under Grants 50 OK 1102, 50 OK
1103 and 50 OK 1104. We are GREATful to the SOFIA operations team for their help and support throughout the course of the observations and after.  The authors would like to express our gratitude to the developers of the many C++ and Python libraries, made available as open-source software, in particular this research has made use of the NumPy \citep{numpy}, SciPy \citep{scipy} and matplotlib \citep{matplotlib} packages.  
\end{acknowledgements}
\bibliographystyle{aa.bst}          
\bibliography{ref.bib}
\begin{appendix}
\section{Sideband deconvolution}\label{appendix:dsb_ceonvolution}

\begin{figure}
    \centering
    \includegraphics[width=0.495\textwidth]{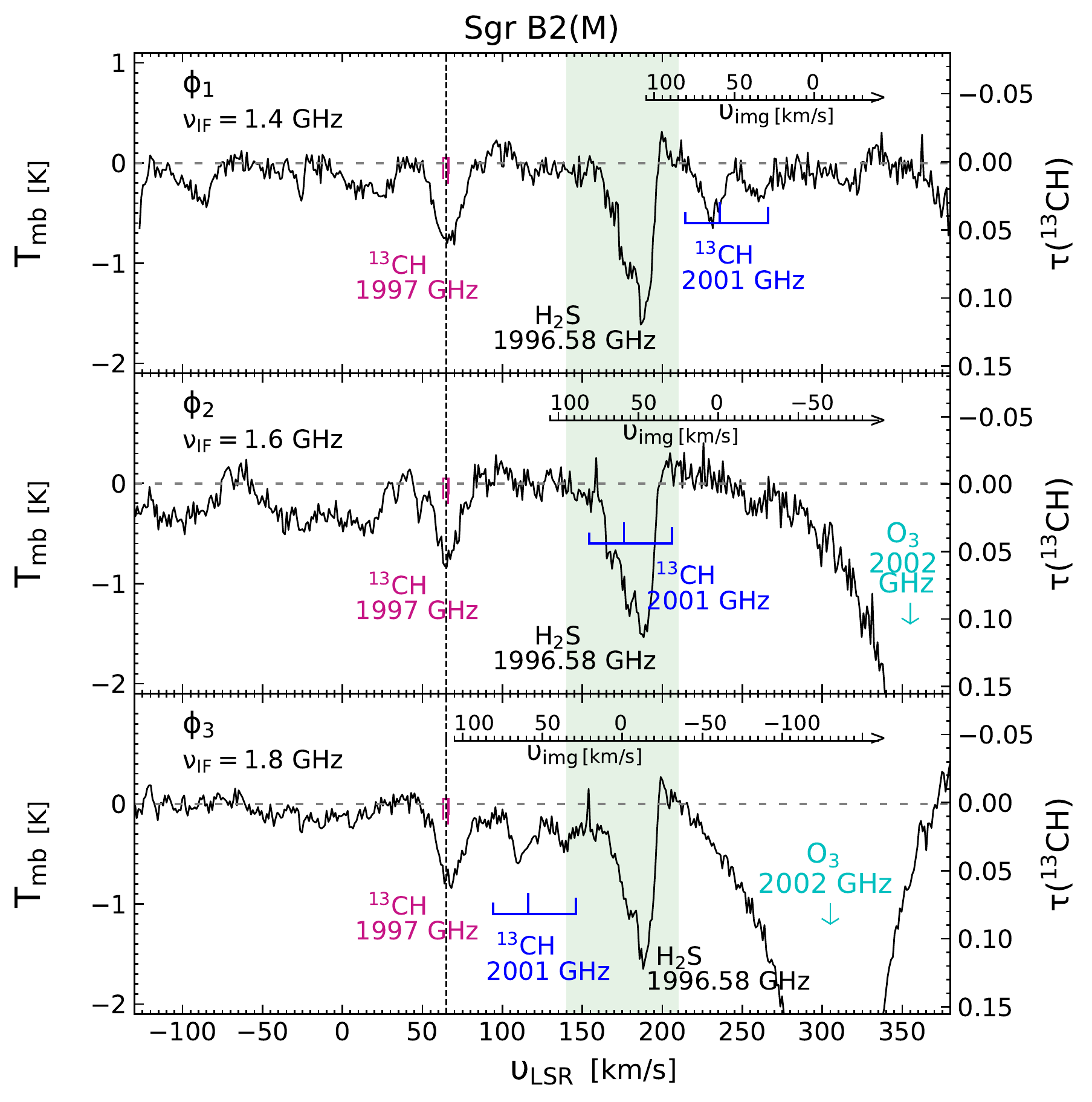}
    \caption{Calibrated and baseline-subtracted observed \ttch spectra at an IF setting of 1.4~GHz (top panel), 1.6~GHz (central panel), and 1.8~GHz (bottom panel) toward Sgr~B2(M). The secondary y-axis in each panel presents the optical depth scale. Signal bandpass contamination, possibly from the $J_{K^{-},K^{+}} = 4_{3,2} \rightarrow 3_{2,1}$ transitions of H$_{2}$S at 1996.589~GHz is displayed by the green shaded region. The hyperfine transitions corresponding to the 1997~GHz $\Lambda$-doublet arising from the signal band are displayed in pink while that of the 2001~GHz $\Lambda$-doublet arising from the image band are displayed in blue. The change in IF pushes both the second $\Lambda$-doublet as well as the atmospheric ozone feature at 2002.347 GHz (telluric rest frame, marked in cyan) towards the signal band features. The secondary x-axis in each panel represents the image band LSR velocity scale.}
    \label{fig:IF_settings_contamintion}
\end{figure}

\begin{figure}
    \centering
    \includegraphics[width=0.495\textwidth]{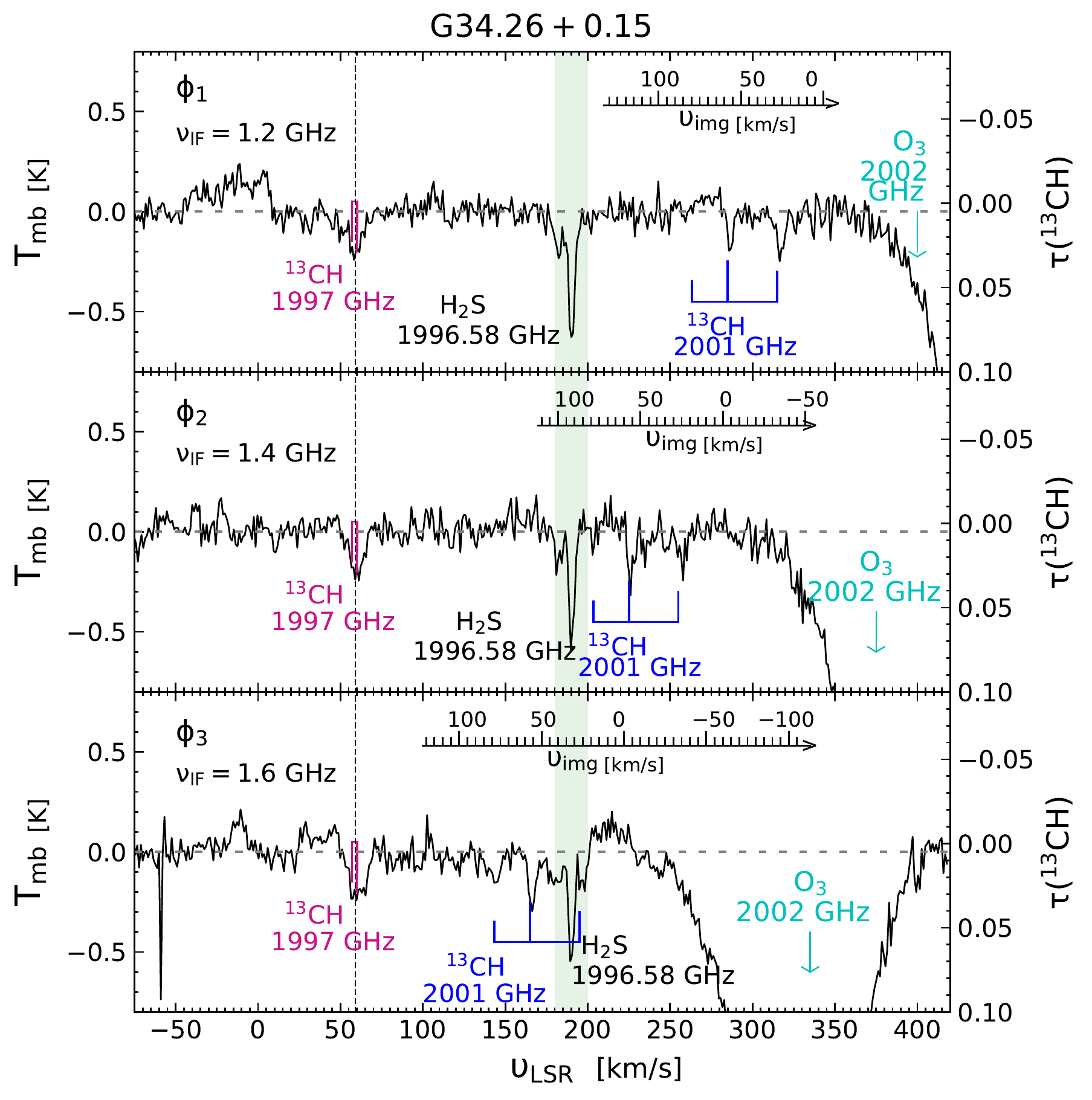}
    \caption{Same as Fig.~\ref{fig:IF_settings_contamintion}, but toward G34.26+0.15 for IF settings of 1.2, 1.4, and 1.6~GHz (from top to bottom).}
    \label{fig:IF_settings_contamintion_G34P26}
\end{figure}

\begin{figure}
    \centering
    \includegraphics[width=0.495\textwidth]{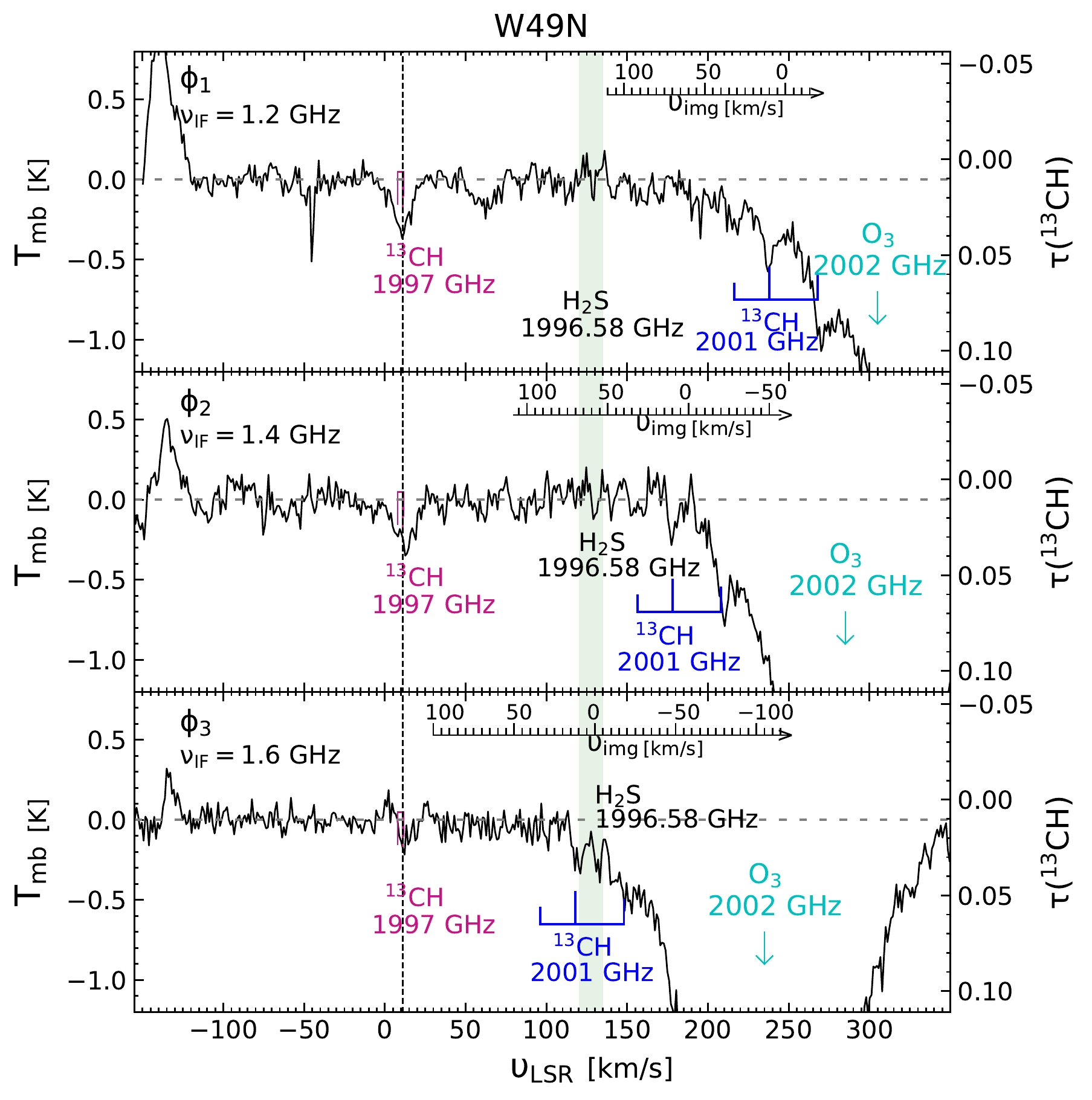}
    \caption{ Same as Fig.~\ref{fig:IF_settings_contamintion}, but toward W49(N) for IF settings of 1.2, 1.4, and 1.6~GHz (from top to bottom) .}
    \label{fig:IF_settings_contamintion_W49N}
\end{figure}

\begin{figure}
    \centering
    \includegraphics[width=0.495\textwidth]{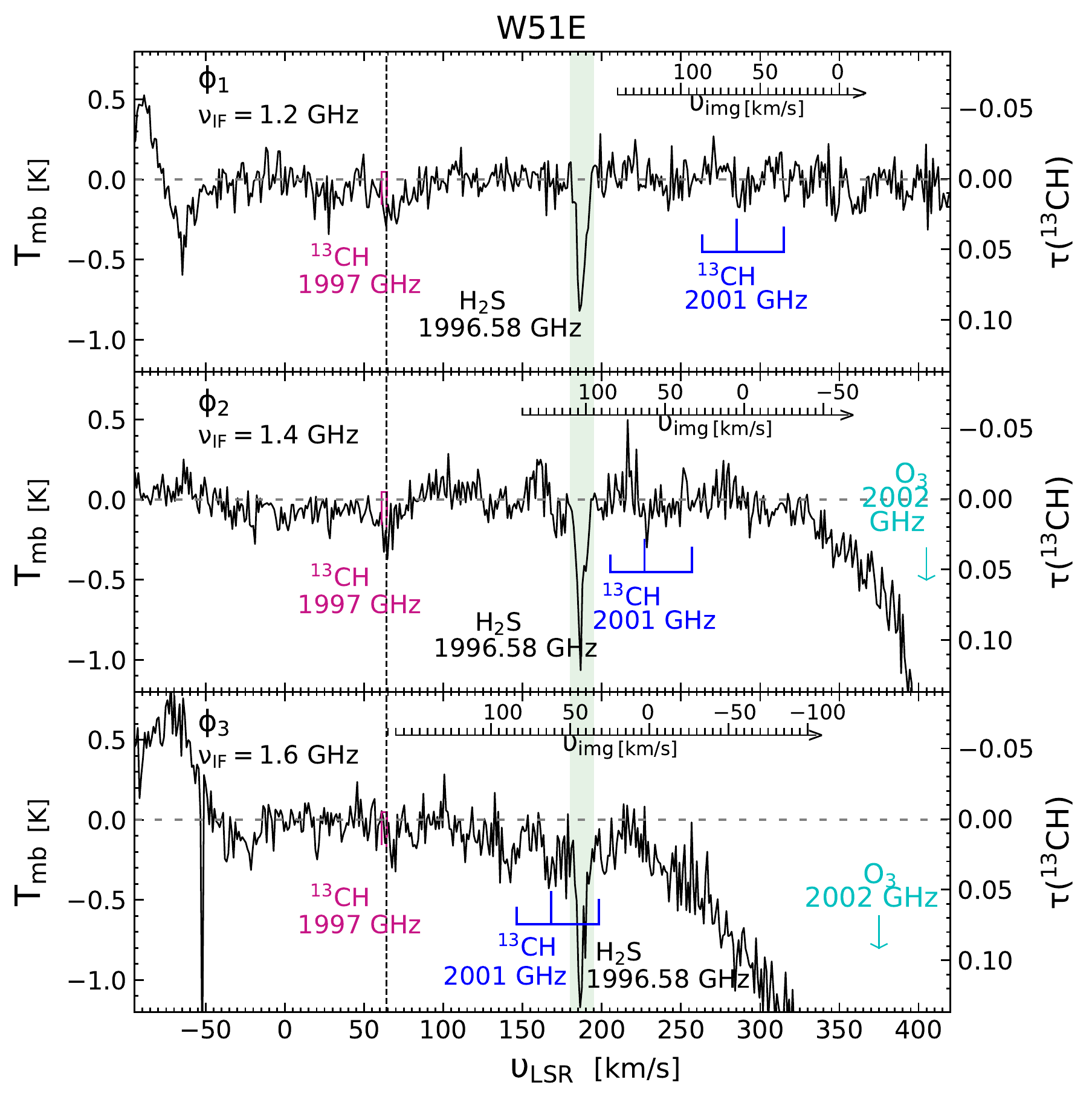}
    \caption{ Same as Fig.~\ref{fig:IF_settings_contamintion}, but toward W51E for IF settings of 1.2, 1.4, and 1.6~GHz (from top to bottom).}
    \label{fig:IF_settings_contamintion_W51E}
\end{figure}

In Fig.~\ref{fig:IF_settings_contamintion} we display the \ttch spectra toward Sgr~B2(M) obtained from the different IF settings at 1.4, 1.6, and 1.8~GHz, respectively. The secondary x-axis displayed in each panel showcases the velocity scale in the image band LSR. Note that each IF offset of 0.2~GHz corresponds to a shift of the image band features by 60~km~s$^{-1}$. When using an IF setting of 1.4~GHz, both sets of $\Lambda$-doublet features are well separated, with the hyperfine components corresponding to the 1997~GHz doublet originating from the signal band while those corresponding to the 2001~GHz doublet arise from the image band. The \ttch (signal band) $\Lambda$-doublet are seen in absorption with a deeper absorption feature seen at 1996.589~GHz possibly arising from the $J_{K^{-},K^{+}} = 4_{3,2} \rightarrow 3_{2,1}$ transitions of H$_{2}$S. Several transitions of H$_{2}$S have previously been detected towards the Sgr B2(M) \citep{tieftrunk1994so} and Sgr B2(N) \citep{neill2014herschel} SFRs, with the higher frequency transitions typically seen in absorption towards the envelope of these hot-cores. This transition lying close to 1996~GHz was not previously detected using HIFI/\textit{Herschel} since it lies outside the tuning range of HIFI Band 7. However, the $J_{K^{-},K^{+}} = 6_{1,6} \rightarrow 5_{0,5}$ high-energy transition of H$_{2}$S at 1846.76~GHz with a lower level energy ${\sim}$239~K was observed using HIFI/\textit{Herschel} toward Sgr B2(M) as a part of the HEXOS guaranteed time key programme\footnote{HIFI/\textit{Herschel} observations of EXtra-Ordinary Sources (HEXOS) \citep{bergin2010herschel} was aimed to investigate the chemical composition of several sources in the Orion and Sgr B2 SFRs.} with observation id 1342206640 showcases a similar absorption profile with comparable line-widths and peak temperatures, which further validates our assumption of this line arising from H$_2$S.

Despite pushing the 2001~GHz $\Lambda$-doublet transitions into the atmospheric ozone feature at 2002.347~GHz, the additional observing setup with an IF of 1.6~GHz was used to confirm the origin of the deeper absorption feature as being from the signal band. Using this setup it is clear that the deeper H$_{2}$S absorption feature must arise from the signal band while further confirming our detection of \ttch absorption toward the envelope of the SFR. However, we require a third IF setting in order to confirm the presence of weak sight-line absorption features. When increasing the IF offset to 1.8~GHz, the \ttch features arising from the image band move closer towards the \ttch features arising from the signal band. We compare the average of the residuals between two independent IF settings with the average of the IF co-added spectrum. Whilst affirming the presence of weaker LOS absorption features by suppressing the image band contributions, this does not present a complete DSB deconvolution. This can be mathematically formulated as follows:\\

Representing the DSB \ttch spectrum obtained from each IF setting as a function of velocity as,
\begin{equation}
   \phi(\upsilon)_{i} = \phi_\text{sig}(\upsilon) + \phi_\text{img}(\upsilon_i - \upsilon) \, ,
   \label{eqn:dsb_spec}
\end{equation} 
where, $i = 1, 2, 3$ corresponding to the different IF settings at 1.4, 1.6, and 1.8~GHz, respectively, $\upsilon_{i}$ is the signal-band velocity at which the image-band velocity is equal to zero and $\phi_\text{sig}$, and $\phi_\text{img}$ represent the signal, and image band contributions. From Eq.~\ref{eqn:dsb_spec} the average of the IF co-added spectrum is,
\begin{equation}
   \phi(\upsilon)_\text{avg} = \phi_\text{sig}(\upsilon) + \frac{1}{3}\left(\phi_\text{img}(\upsilon_{1} - \upsilon) + \phi_\text{img}(\upsilon_{2} - \upsilon) + \phi_\text{img}(\upsilon_{3} - \upsilon) \right) \, ,
   \label{eqn:avg}
\end{equation}
The set of independent residuals between the different IF settings using the spectrum with an IF = 1.4~GHz as reference is,
\begin{eqnarray}
\phi(\upsilon)_\text{1-2} &= \phi(\upsilon)_1 - \phi(\upsilon)_2 \, ,   \\
\phi(\upsilon)_\text{1-3} &= \phi(\upsilon)_1 - \phi(\upsilon)_3  \, .  
\end{eqnarray}
The average of these residuals will reduce the image band features from the spectra with IF settings 1.6 and 1.8~GHz by half.
\begin{align}
\phi_{\text{residual-avg}} &=  \left( \phi(\upsilon)_\text{1-2} + \phi(\upsilon)_\text{1-3} \right)/2 \\
 &= \left( 2\phi_\text{img}(\upsilon_1 - \upsilon) - \phi_\text{img}(\upsilon_2 - \upsilon) - \phi_\text{img}(\upsilon_3 - \upsilon) \right)/2
\label{eqn:residual-avg}
\end{align}
Subtracting Eq.~\ref{eqn:residual-avg} from Eq.~\ref{eqn:avg} yields,
\begin{align}
    \phi_\text{resultant} &= \phi_\text{avg} - \phi_\text{residual-avg} \\
\begin{split}
    &= \phi_\text{sig}(\upsilon) -\frac{2}{3}\phi_\text{img}(\upsilon_{1} - \upsilon) +  \frac{5}{6}\phi_\text{img}(\upsilon_2 - \upsilon) \\ &+ \frac{5}{6}\phi_\text{img}(\upsilon_3 - \upsilon) 
    \end{split}
\end{align}
While this exercise models the contribution of the 2001~GHz transitions of \ttch, it does not confirm the profiles of the LOS features. Additionally, discrepancies in the noise level between the three spectral tunings arising from different integration times between them may lead to errors in the sideband reconstruction. However, the detection of $^{13}$CH, at the systemic velocity of the Sgr B2(M) molecular cloud, remains undeterred. 

A similar DSB deconvolution was carried out for G34.26+0.15, W49(N) and W51E using a nominal IF setting of 1.2, 1.2, and 1.4~GHz, respectively as reference. The spectra observed using each of the three different IF tunings toward each of these sources are displayed in Figures~\ref{fig:IF_settings_contamintion_G34P26}, ~\ref{fig:IF_settings_contamintion_W49N} and \ref{fig:IF_settings_contamintion_W51E}, respectively. We clearly detect the \ttch 1997~GHz $\Lambda$-doublet component at the respective systemic velocities of G34.26+015, W49(N), and W51E, in all three IF settings at 1.2, 1.4 and 1.6~GHz, however the hyperfine lines of the 2001~GHz $\Lambda$-doublet component either lie at the edge of the ozone feature for an IF setting of 1.2~GHz or move into the H$_{2}$S absorption feature for the higher IF tuning at 1.6~GHz. Despite, confirming the detection of \ttch toward the systemic velocities of these sources, it is once again difficult to establish the presence of any LOS features post the sideband deconvolution.

\section{\texorpdfstring{\ttch}{13CH}  spectrum toward W3(OH)}\label{appendix:w3oh_spec}

In this Appendix we display the \ttch spectrum toward W3(OH). 
\begin{figure}
    \centering
    \includegraphics[width=0.45\textwidth]{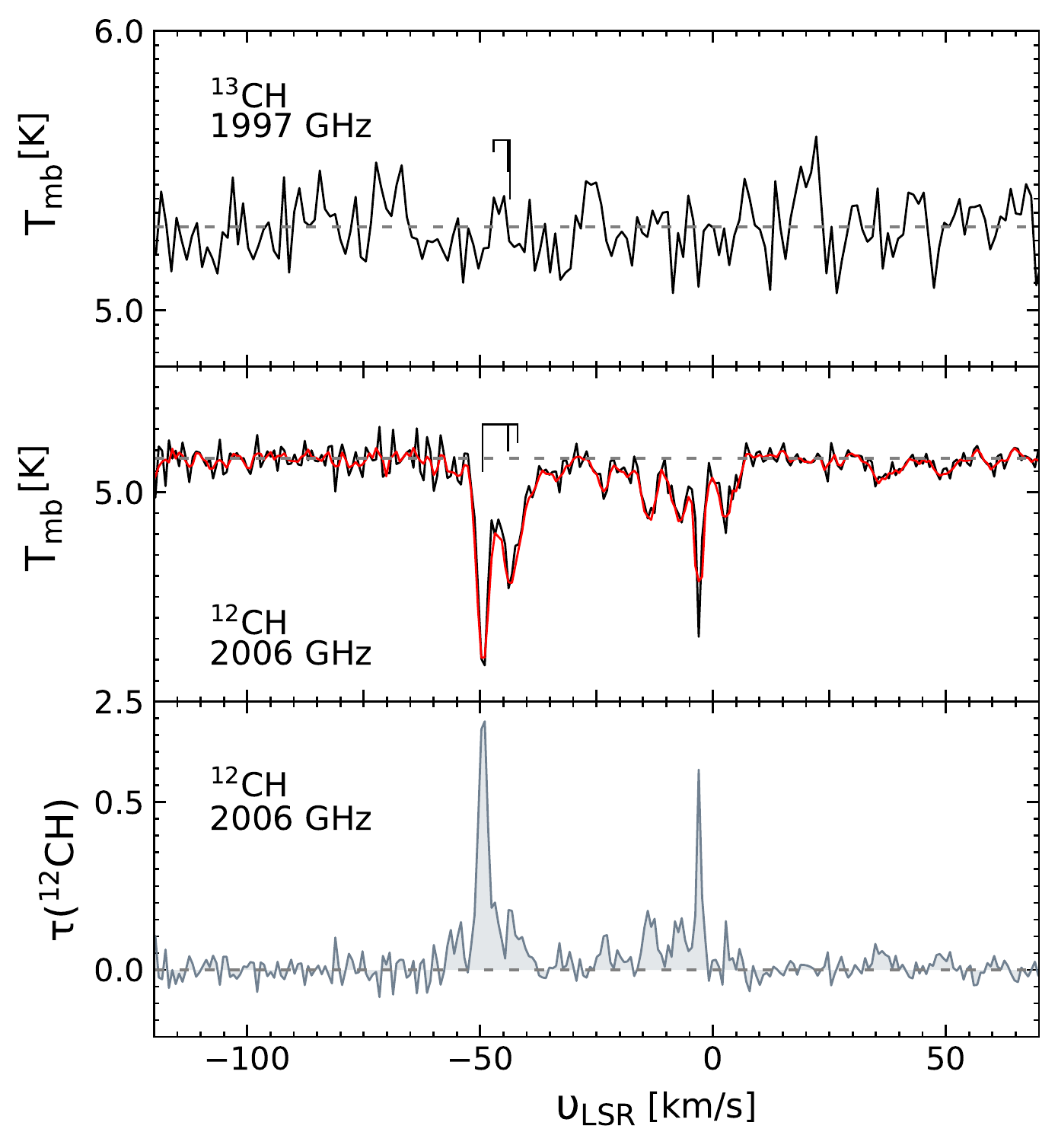}
    \caption{Spectra of the $N, J=2,3/2\rightarrow1,1/2$ transitions of \ttch near 1997~GHz (top) and CH near 2006~GHz (middle) observed towards W3(OH) in black. The Wiener filter model of the observed CH spectrum is overlaid in red. The bottom panel presents the Wiener filter deconvolved CH spectrum in optical depth scales.}
    \label{fig:w3oh_spec}
\end{figure}

\end{appendix}

\end{document}